\documentclass{article}
\usepackage[T1]{fontenc}
\usepackage{graphicx}
\usepackage{epsf,rotate}
\usepackage{epsfig}
\usepackage{rotating}
\usepackage{cite}

\makeatletter

\newcommand{\LyX}{L\kern-.1667em\lower.25em\hbox{Y}\kern-.125emX\spacefactor1000}

\usepackage[T1]{fontenc}

\makeatletter

\usepackage[T1]{fontenc}
\usepackage{geometry}


\makeatother

\begin{document}

\title{The Damage Spreading Method in Monte Carlo Simulations: A brief overview and applications to confined magnetic materials}
\author{M. Leticia Rubio Puzzo and Ezequiel V. Albano}
\date{Instituto de Investigaciones Fisicoqu\'{\i}micas Te\'{o}ricas
y Aplicadas (INIFTA), UNLP, CONICET,
Casilla de Correo 16 Sucursal 4, (1900) La Plata, Argentina.}
\maketitle

\begin{abstract}

The Damage Spreading (DS) method allows the investigation of the effect
caused by tiny perturbations, in the initial conditions of physical systems,
on their final stationary or equilibrium states. The damage $(D(t))$ is determined during
the dynamic evolution of a physical system and measures the time dependence
of the difference between a reference (unperturbed) configuration and an
initially perturbed one.
In this paper we first give a brief
overview of Monte Carlo simulation results obtained by applying the DS method.
Different model systems under study often exhibit a transition between a state
where the damage becomes healed (the frozen phase) and a regime where
the damage spreads arriving at a finite (stationary) value (the damaged phase),
when a control parameter is finely tuned.
These kinds of transitions are actually true irreversible phase transitions themselves,
and the issue of their universality class is also discussed.

Subsequently, the attention is focused on the propagation of
damage in magnetic systems placed in confined geometries. The
influence of interfaces between magnetic domains of different
orientation on the spreading of the perturbation is also
discussed, showing that the presence of interfaces enhances the
propagation of the damage. Furthermore, the critical transition
between propagation and nonpropagation of the damage is discussed.
The results analyzed indicate that, in some cases as in the
Abraham's Model and in the standard Ising magnet (Glauber
dynamics), there is clear evidence showing that the DS transition
and the critical transition of the physical system (in these cases
the wetting and the ferromagnet-paramagnet transitions,
respectively) occur at different critical points. However, in the
case of the corner geometry, the critical points of both
transitions$-$damage spreading and cornerfilling$-$coincide within
error bars. It is found that, at criticality, the damage obeys a
power-law behavior of the form $D(t) \propto t^{\eta}$, where
$\eta$ is the damage spreading critical exponent. The evaluation
of critical exponents allows the identification of three
propagation regimes: i) inside the magnetic domains the
propagation is slow $(\eta = 0.40(2))$, ii) the fast propagation
is observed along the interface between domains $(\eta =
0.90(2))$, iii) the alternating propagation across interfaces and
inside domains is consistent with an exponent lying between the
previous cases, namely $\eta = 0.47(1)$. In all cases, the
determined critical exponents suggest that the DS transition does
not belong to the universality class of Directed Percolation,
unlike many other systems exhibiting irreversible phase
transitions. This result reflects the dramatic influence of
interfaces on the propagation of perturbations in magnetic
systems.

~


~


\end{abstract}

\pagebreak

\section{Introduction.}

One of the most interesting challenges in the theory of dynamic systems is the
understanding of the dependence of the time evolution of physical observables
on the initial conditions, because very often a small perturbation in the initial parameters could completely change their behavior \cite{hinrich2000}.
Within this context, it is interesting to study the time evolution of such
perturbations in order to investigate under which conditions a small initial
perturbation may grow up indefinitely or, eventually, it may vanish and become healed.

In order to understand this behavior, Kauffman introduced the concept of
Damage Spreading (DS) \cite{kauf}. In order to implement the DS method in
computational simulations \cite{herr90, binderMC}, two configurations or
samples $S$ and $S'$, of a certain stochastic model, are allowed to evolve
simultaneously. Initially, both samples differ only in the state of a small
number of sites. Then, the difference between $S$ and $S'$ can be considered
as a small initial perturbation or damage.

The time evolution of the perturbation can be followed by evaluating the total
damage or ``Hamming distance'' defined as

\begin{equation}
D(t)= \frac{1}{N} \sum_{i} D_i (t)=\frac{1}{N} \sum_{i} 1 - \delta_{S_i (t),S'_i (t)} ,
\label{defdam}
\end{equation}

\noindent where $D_i(t)$ is the damage of the site labeled with the index $i$ at
time $t$, $\delta_{S_i (t), S'_i (t)}$ is the delta function and the summations
in equation (\ref{defdam}) run over the total number of sites of the system $N$.

By starting from a vanishing small perturbation $D(t=0)\rightarrow 0$, one may
expect at least two main scenarios, namely: a) $D(t\rightarrow \infty) \rightarrow 0$
and the perturbation is irrelevant because the damage heals;
or b) $D(t\rightarrow \infty)$ assumes some non-zero value, and the damage spreads.
These two situations indicate a weak or a strong influence of the initial
conditions, respectively \cite{bind1,herr}. Of course, an intermediate or marginal case where the number of damaged sites approaches a finite positive number, which does not increase proportionally to $N$, may also occur. 
In this case, after normalization by $N$ one has that the equation (\ref{defdam}) gives $D(t\rightarrow \infty) \equiv 0$, but the damage has not healed out.
 When the system arrives either
at the state of zero damage or complete damage ($D=0$ or $D=1$, respectively),
it will remain in this situation indefinitely.
For this reason, a transition between these two states is irreversible and
could be related to Directed Percolation processes \cite{percol},
for which an irreversible critical transition occurs from an active to
an inactive state (absorptive state).
In those models belonging to the
universality class of directed percolation, when the system is trapped in an
absorptive state, it is then impossible to recover the activity by changing
the control parameter.

The first studies of DS in physical systems were applied to the Ising model,
spin glasses and the Kauffman cellular automata, and they appeared in the
mid-eighties \cite{martin, derrS, stan, derrW}.
Subsequently, this technique
has also been applied to the study of several different models such as
Ising models \cite{derrS,stan,derrW,coni,cost,leca,mariz89,glot1,mariz,poole90,
glot2,nobre92,glot,batrouni,zhang,glotzer93,hunter93,stau,matz,grah,tama,gropen,gras1,rojd,gras2,gras3,gropen95,mont,wang,moreira,lima,hinrich97,vojt1,vojt2,vojt3, junior,vojt4,vojt5,neves,hinrich98,argo,gleiser,argo00,liu,rubio2001,nobre01,rubio2002, rubio2002PRB,tome,rubio2005,rubio2007},
spin glasses \cite{derrW,arcang89,arcang89b,dacruz,boissin,campbell91,campbell93,campbell94,almeida,wang96,wappler,heer},
Potts models with q-states \cite{mariz90,bibi,silv,redinz98,souz,redinz01,anjos},
the Heisenberg model \cite{miranda,costa00},
the XY model \cite{golinelli,chiu},
a discrete ferromagnet \cite{mariz93},
two-dimensional trivalent cellular structures \cite{guo1,guo2},
biological evolution \cite{staufer94,vale}, and
cellular automata \cite{herr,bagnoli92,tsal,tsal94,tome94,hinrich96,mone,bagnoli}.
Also, non-equilibrium systems \cite{kauf,grah,mone,miranda89,albanoZGB,albanoRDif,
bhowal,odor,gleiser01}, SOS models \cite{kim1,kim2,kim3},
opinion dynamics \cite{fortunato,klietsch} and
small world networks \cite{svenson,medeiros},
have been characterized by means of the DS method.

In view of the great interest attracted by this field of research of interdisciplinary
application, the aim of this work is to present a brief overview of the state of the
art in studies of the Damage Spreading transition, focusing our attention on recent
results obtained for magnetic systems in confined geometries, which are helpful in understanding the propagation of perturbations in nano- and micromaterials.

The manuscript is organized as follows: in Section 2 we briefly describe the
archetypical models used in the study of DS, namely the Domany-Kinzel cellular
automata and the Ising model. 
In Section 3 we describe the DS method in these basic models, and Section 4 is devoted to the discussion on the main characteristics of DS. 
In Section 5, in first place we present a brief discussion of the equilibrium
configurations that can be found for Ising systems in confined geometries, and therefore we discuss the results obtained for DS in these systems.
Finally, our conclusions are stated in Section 6.

\section{Definition of basic models.}

Both, the Domany-Kinzel (DK) cellular automata model and the Ising magnet have
become archetypical systems for the study of DS.
So, for the sake of completeness we give brief descriptions of both models.

\subsection{The Domany-Kinzel Cellular Automata Model.}

The DK model \cite{domkin,kinzel} is a family of the $(1+1)$ dimensional stochastic
cellular automata with two parameters, $p_1$ and $p_2$, which simulate the time
evolution of interacting active elements in a random medium.
The DK model consists of a linear chain of $N$ sites ($S_i$) that can take two
possible values, usually $0$ and $1$ (empty and occupied sites, respectively).
The state of each site $i$ at time $t+1$ [$S_i(t+1)$] depends only upon the state
at time $t$ of the two nearest neighbors [$S_{i-1}(t)$ and $S_{i+1}(t)$],
according to the transition probability [$P[S_i(t+1)\mid S_{i-1}(t),S_{i+1}(t)]$]
defined as

\begin{eqnarray}
P[1\mid 0,0]&=&0 \\ \nonumber
P[1\mid 0,1]&=& P[1 \mid 1,0]= p_1 \\ \nonumber
P[1\mid 1,1]&=& p_2 \quad ,
\label{probDK}
\end{eqnarray}

\noindent where $P[0|\bullet, \bullet]=1- P[1|\bullet, \bullet]$ and the
parameters $p_1$ and $p_2$ represent the probabilities that the site $i$
is occupied if exactly one or both of its neighbors are also occupied,
respectively.

Domany and Kinzel demonstrated \cite{domkin,kinzel} the existence of two phases,
depending on the values of the parameters $p_1$ and $p_2$, a frozen and an active
phase, separated by a critical line. In the active phase, there exists a stationary
state that it is governed by fluctuations, while in the frozen phase all initial
states lead to an absorbing state. There is strong numerical evidence that this
phase transition belongs to the universality class of Directed Percolation
(except for its upper terminal point).
The exceptional behavior at the upper
terminal point of the critical line is due to an additional symmetry between
active and inactive sites along the line $p_2=1$.
Here, the DK model has two
symmetric absorbing states given by the empty and the fully occupied lattices,
respectively.

\subsection{The Ising Model.}

The other archetypical system used to study the DS transition is the
Ising model \cite{ising}. In this case, each site of the lattice represents
a spin variable. In the ferromagnetic case, the spins have an energetic
preference to adopt the same direction. The Hamiltonian of this system can be written as

\begin{equation}
{\cal{H}}=-J\cdot \sum _{<i,j>}\sigma _{i}\sigma _{j} - H \sum _{<i,j>}\sigma _{i}
\label{hamis}
\end{equation}

\noindent where $\sigma _{i}$ is the Ising spin variable that can assume two
different values $\sigma_i = \pm 1$, the indexes $1\leq i,j \leq N$ are used
to label the spins, $J>0$ is the coupling constant of the ferromagnet, $H$ is
the external magnetic field, and the summation runs over all the nearest-neighbor pairs of spins.
In the absence of an external magnetic field ($H=0$) and at low temperature,
the system is, for more than one dimension, in the ferromagnetic phase and, on average, most spins are pointing
in the same direction. 
In contrast, at high temperature the system maximizes the entropy,
thermal fluctuations break the order and the system is in the paramagnetic phase.
This ferromagnetic-paramagnetic critical transition is a second-order phase transition
and it occurs at a well-defined critical Temperature ($T_C$). In the two-dimensional
case, one has exactly $k T_C/J = 2/ln(1+\sqrt{2})=2.269....$, where $k$ is the Boltzmann constant.

\section{Damage Spreading in the basic models.}

\subsection{Damage Spreading in the DK Model.}

One of the most interesting behaviors of DS was first discovered in
the $(1+1)$-dimensional DK cellular automata \cite{kinzel, domkin}. 
In fact, in addition to the two known phases (frozen and active) of
the phase diagram of the DK model, already discussed in the
previous section, Martins et al. \cite{mart} found a third phase
related to the spreading of the damage. 
This ``new'' phase, which
lies in the active phase, simultaneously exhibits regions where
the damage spread and heals, and it is extremely sensitivity to
the initial conditions. Subsequently, other authors
\cite{gras3,zebende,martzeb,rieger} determined the boundary of
this phase more precisely. Independently, Mean Field
approximations applied to different systems
\cite{bagnoli,rieger,kohring,tome94} confirmed the existence of
this ``chaotic phase'' in the sense that the final state of the
system is sensitive to the initial conditions. 
However, it has
also been realized that the results obtained
\cite{bagnoli,kohring} may depend on the dynamic rules used in the
implementation of the algorithm. 
This characteristic of DS will be explained in detail in next Section.

\subsection{Damage Spreading in the Ising Model.}

In the case of the Ising model, the definition given in equation
(\ref{defdam}) can be rewritten as

\begin{equation}
D(t)=\frac{1}{2N}\sum ^{N}_{l}\left| S^{A}_{l}(t,T)-S^{B}_{l}(t,T)\right|,
\label{eq:dam}
\end{equation}

\noindent where the summation runs over the total number of spins
$N$, and the index $l\,(1\leq l\leq N)$ is the label that
identifies the spins of the configurations. $S^{A}(t,T)$ is an
equilibrium configuration of the system at temperature $T$ and
time $t$, while $S^{B}(t,T)$ is the perturbed configuration that
is obtained from the previous one, at $t=0$, just by flipping few
spins \cite{herr90}. Physically, the definition given by equation
(\ref{eq:dam}) represents the total fraction of spins that are
different in both configurations. Then, one is interested in
investigating under which conditions a small initial perturbation
will grow up indefinitely or eventually will vanish and become
healed.

Notice that by performing Monte Carlo simulations, the same
sequence of random numbers has to be used for both copies, in
order to assure identical realizations of thermal noise.
Furthermore, to study the time evolution of the DS, a meaningful
definition of the Monte Carlo time step (mcs) is necessary. For
this purpose, the standard definition is adopted according to that
during one mcs all $N-$spins of the sample are flipped once, on
the average.

\section{Main characteristics of the Damage Spreading.}

\subsection{On the Dependence of Damage Spreading with the Simulation Algorithm.}

As it was mentioned previously, the DS behavior depends on the dynamic rules used to implement the algorithm, in Monte Carlo Simulations.
In order to understand this dependence, it is very useful to remind
the reader that the detailed balance condition
\cite{hinrich2000,binderMC} assures that the system will arrive at
an equilibrium state, but it does not establish the way of this
evolution. In other words, detailed balance does not univocally
determine the dynamic rules that have to be applied to go from a
given configuration to the next. Therefore, this situation opens
the possibility on choosing different transition probabilities
between states, which implies that different dynamic rules can be
applied to the same physical system. In the case of the Ising
Model, the equilibrium state can be generated by different dynamic
rules, e. g. Heath Bath, Glauber, Kawasaki and Metropolis
dynamics, which represent different dynamics that allow the system
to arrive at an equilibrium state.

In principle, it was expected that DS would not depend on the
intrinsic dynamics of the simulation, and one could find regular
and chaotic phases that could be identified with the properties of
the equilibrium system \cite{stan,derrW}. However, subsequent
studies showed that different dynamic rules implies the occurrence
of different behavior in the DS, such as Glauber versus Metropolis
\cite{mariz,jan}, Q2R \cite{glot1} or Kawasaki \cite{glot2}.
Furthermore, depending on the type of updates, this means in which
way the sites of the lattice are chosen (randomly, typewriter,
chessboard, etc.), the results obtained for DS and the critical
temperature of the transition between propagation and
nonpropagation of the damage are different \cite{nobre92,
glotzer93,vojt5}.

Summing up, the observed behavior is not surprising since DS is a
dynamic process, and for this reason, it is reasonable to expect
that it may depend on the dynamic rules applied. As an example, it
is useful to consider the Ising model in two dimensions and
compare the results obtained using both Glauber \cite{stan} and
Heath Bath \cite{derrW} dynamics. For the Ising model simulated by
using the Glauber dynamics, Stanley et al. \cite{stan} and Mariz
et al. \cite{mariz} found that in the paramagnetic phase ($T >
T_C$, where $T_C$ is the Onsager critical temperature ($T_C =
2.269.....J$)) the system is chaotic ($D(t\rightarrow \infty)$
tends to a finite non-zero value), while in the ferromagnetic
phase the damage heals ($D(t\rightarrow \infty)\rightarrow 0$). On
the other hand, Derrida et al. \cite{derrW} studied DS in the
Ising Model with the Heath Bath dynamics. They found, in contrast
to the case of Glauber dynamics, that for both the paramagnetic
and ferromagnetic phases, damage heals in the limit
$D(0)\rightarrow 0$.

On the other hand, the results obtained by U. Costa \cite{cost}
and Le Ca\"{e}r \cite{leca} in three dimensions and using the
Glauber dynamics, show a behavior similar to the 2D case, but the
critical temperature for the DS transition ($T_D$) is lower than
$T_C$.

\subsection{On the Universality Class of the Damage Spreading Transition.}

The universality class of the continuous and irreversible critical
transition between propagation and non-propagation of damage is
still an open question. Grassberger \cite{gras3} conjectures that
the DS transition may belong to the Directed Percolation
universality class (DP) \cite{pd5,jensen1,voigt} if its critical
point does not coincide with a critical transition of the physical
system, e.g. the critical temperature of the Ising magnet. In the
same paper, Grassberger \cite{gras3} presents a Monte Carlo
simulation study of DS in the DK cellular automata \cite{DK} as a
test for his conjecture. He founds that critical exponents of the
DS transition coincide, within error bars, with the exponents of
the DP universality class, both in two and three dimensions.

An analytical justification for Grassberger's conjecture is given
by Mean Field studies reported by Bagnoli \cite{bagnoli} and the
subsequent exact results obtained by Kohring and Schreckenber
\cite{kohring}. They found that, for certain limits of the
transition probability between states, the dynamics of DS in the
DK cellular automata is identical to the evolution of the DK
itself, and for this reason, the DS transition belongs to the DP
universality class. These results were later extended to other
regions of the phase diagram of the DK Model \cite{hinrich96}.

Numerical simulations of different models also showed that the DS
transition is characterized by critical exponents of the DP
universality class, as in the case of the 2D Ising Model with
Swendsen-Wang dynamics \cite{hinrich98}, as well as in a
deterministic cellular automata with small noise \cite{bagnoli92}.

However, there are also other systems where the DS transition has a non-DP behavior,
such as the case of the Kauffman Model \cite{kauf}. In general, it is expected that
if the order parameter has a $Z_2$ symmetry, the DS transition may not belong to
the DP class \cite{hinrich97}.

In the Ising model with Glauber dynamics, Stanley et al.
\cite{stan} determined that the DS transition and the
paramagnet-ferromagnet transition coincide. For this reason, it is
expected that the DS transition may not belong to the DP
universality class.

In three dimensions, Costa \cite{cost} and Le Ca\"{e}r \cite{leca}
found that $T_D/T_C \approx 0.96$ and $T_D/T_C \approx 0.91$,
respectively. A few years later, Grassberger \cite{gras2}
determined more precisely the critical temperature for the DS
transition, and he found that $T_D = 0.992(2) T_C$ in two
dimensions and $T_D = 0.9225(5) T_C$ in three dimensions. He also
determined the value of the critical exponent $\delta$ governing
the time dependence of the survival probability of damage, showing
that it coincides, within error bars, with the accepted value
corresponding to the DP universality class.

On the other hand, Vojta \cite{vojt1,vojt3}, studying the dynamic
stability of the kinetic Ising model with Glauber dynamics and
using a Mean Field approximation, found that there exists a
critical temperature for the Damage propagation-nonpropagation
transition given by $T_D \simeq 1.739J \simeq 0.826T_C$. Also,
Vojta generalized these results in the presence of an external
magnetic field $h$. He found that $h$ causes an increase of the
critical temperature $T_D$ and stabilizes the non-chaotic phase.
The general behavior of $T_D$ as a function of $h$ is given by

\begin{equation}
\frac{T_D(h)}{J} = \frac{1}{1-h/J} \quad \mbox{in the limit $h/J \rightarrow 1$.}
\end{equation}

DS has also been studied in the Ziff-Gulari-Barshad (ZGB) model
\cite{zif}, for the catalytic oxidation of carbon monoxide. The
ZGB model exhibits a second-order irreversible phase transition
between an active state with production of $CO_{2}$ and a poisoned
(absortive) regime \cite{zif} where the reaction stops
irreversibly, which is known to belong to the universality class
of DP \cite{loscar}. By performing Monte Carlo simulations of DS
in the ZGB model in 2D, Albano \cite{albanoZGB} showed that there
exist both chaotic and regular phases, and that the transition
between them lies within the reactive phase. So, the DS transition
is not coincident with the model transition that takes place
between a poisoned and a reactive phase. However, the reported
critical exponent $\delta \simeq 0.65(2)$ is quite different from
that corresponding to the DP universality class, namely $\delta =
0.451$ \cite{voigt}. So, the question about the universality class
of the DS transition in the ZGB model is still open.

Another interesting possibility is to study the interplay between
reversible (equilibrium) critical transitions of confined systems
and the irreversible DS transition. In particular, for the case of
magnetic systems confined between rigid walls, it is interesting
to study the relationship between on the one hand, the DS
transition and on the other hand, the paramagnet-ferromagnet, the
wetting and the corner-filling transitions. In the last examples
(wetting and corner transitions), the presence of external
magnetic fields applied to the walls of the system promotes the
presence of interfaces between magnetic domains of different
orientation, and therefore it is interesting to study the effect
of these interfaces on the propagation of damage. In fact, it have
been observed  \cite{rubio2001,rubio2002,rubio2002PRB,rubio2007}
that the presence of interfaces enhances the spatiotemporal
spreading of damage.

\section{Damage Spreading in the Confined Ising Model.}

Here, we focused our attention to the study of DS in the Ising Model when the ferromagnet is confined between walls that exhert surface magnetic fields.
For this purpose, first it is worth studying the main properties of equilibrium configurations, since they are the starting point for the study of DS.
Subsequently, the study of DS is actually addressed.

\subsection{Equilibrium Configurations of the Ising Magnet in Confined Geometries.}

\subsubsection{Strip Geometries and the Wetting Transition.}

The study of the properties of thin$-$confined$-$films has attracted
growing attention in the last decades not only due to the interest
in the understanding of the properties of confined and low
dimensional materials, but also to the existence of many potential
applications in the fields of nanoscience and nanotechnology.

A thin film can be modeled by using the Ising magnet in a confined
geometry. So, if one has a film in a stripped geometry of size $L
\times M$ ($L\ll M$), the Hamiltonian of the Ising magnet given by
equation (\ref{hamis}) can be redefined as

\begin{equation}
{\cal H}= -J\sum ^{M,L}_{<ij,mn>}\sigma _{ij}\sigma _{mn} -
h_{1}\sum ^{M}_{i=1}\sigma _{i1} - h_{L}\sum ^{M}_{i=1}\sigma _{iL}
\label{eq:ham}
\end{equation}

\noindent where $\sigma_{ij}$ is the Ising spin variable
corresponding to the site of coordinates $(i,j)$, $J>0$ is the
coupling constant of the ferromagnet and the first summation of
(\ref{eq:ham}) runs over all nearest-neighbor pairs of spins such
as $1\leq i\leq M$ and $1\leq j\leq L$. The second (third)
summation corresponds to the interaction of the spins placed at
the surface layer $j=1$ ($j=L$) of the film where the surface
magnetic field $h_{1}$ ($h_{L}$) acts. Such fields are measured in
units of the coupling constant J. Of course, for non-vanishing
surface fields one has to assume open boundary conditions (OBC)
along the $M-$direction of the film. However, for $h_{1}=h_{L}=0$
both OBC and periodic boundary conditions (PBC) may be used. Also,
along the $L-$direction of the strip, OBC are always used.

\begin{figure}
\centerline{{\epsfysize=2in \epsffile{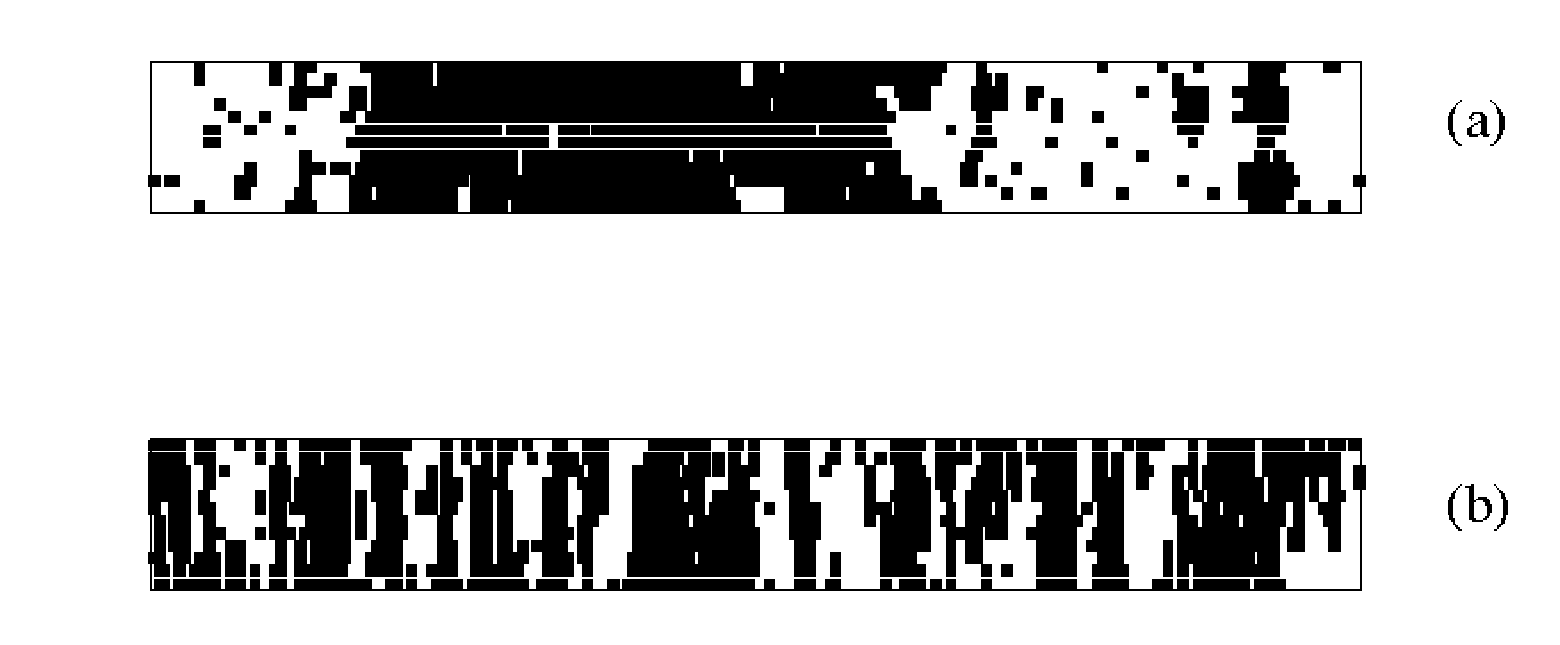}}} 
\caption{Snapshot
configurations corresponding to the Ising magnet as obtained for
$L = 12$, $M = 1200$, $ t = 10000$ mcs, $T = 0.98 T_{C}$, and
using (a) periodic boundary conditions, (b) open boundary
conditions. Note that the horizontal coordinate has been reduced
by a factor of five in comparison with the vertical coordinate,
for the sake of clarity of the picture. Sites taken by down spins
are shown in black while up spins are left white \cite{rubio2001}.
More details in the text.} 
\label{fig1intro}
\end{figure}

Considering PBC connecting the upper and lower surfaces of the
film and neglecting surface fields in the Hamiltonian of equation
(\ref{eq:ham}), one finds standard configurations of the Ising
magnet, as is shown in Figure \ref{fig1intro} (a). In this case,
for $T=0.98T_{C}$ one has rather homogeneous configurations
essentially showing large monodomain structures. On the other
hand, by assuming OBC and keeping $h_{1}=h_{L}=0$, a quite
distinct behavior is observed, as is shown in Figure
\ref{fig1intro} (b), which was also obtained at the same
temperature as in Figure \ref{fig1intro} (a). In fact, near the
bulk critical temperature, the system shows a
quasi-phase transition from a state that is ordered at scales $\xi
< L$ (note that $\xi$ is the standard correlation length) below
$T_C$ to a state that is essentially ordered in the direction
perpendicular to the open boundary, but is disordered in the other
direction. In fact, for $T \leq T_{C}$ (but close to $T_C$) the
system is broken up in a sequence of magnetic domains with spins
of opposite sign. The successive domain walls occur essentially at
random (see e.g. Figure \ref{fig1intro} (b)). It should be noticed
that this particular kind of configuration is the macroscopic
manifestation of the mixing neighbor effect undergone by spins at
the surfaces of the film with open boundaries. For a detailed
discussion on this kind of configuration, see also Albano et al.
\cite{albano1989}.

\begin{figure}
\centerline{{\epsfysize=2in \epsffile{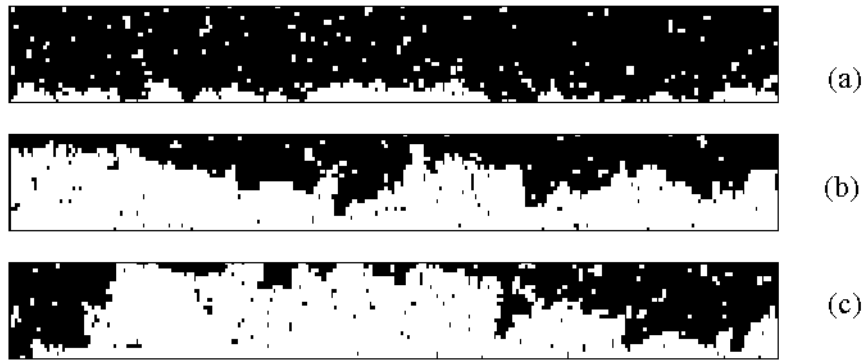}}} 
\caption{Snapshot
configurations obtained after $10^4$ mcs in confined geometries
when short-range fields with opposite signs are applied to the
bottom and top sides of a lattice of size $L=24$ and  $M=1200$. In
these cases, the snapshots are taken at $T=0.80\, T_{C}$ and
different surface fields: (a) $h = 0.4$, within the non-wet phase;
(b) $h = 0.6$, near the critical wetting curve, and (c) $h = 0.8$,
within the wet phase. Note that the horizontal coordinate has been
reduced by a factor of five in comparison with the vertical one,
for the sake of clarity of the picture. Sites taken by down spins
are shown in black while up spins are left white
\cite{rubio2002PRB}. More details in the text.}
\label{fig2prb}
\end{figure}

Also, by considering OBC and competing surface magnetic fields
$h_{1} = -h_{L}$ (see equation (\ref{eq:ham})) another very
interesting scenario takes place. In fact, for $T < T_{C}$, these
competing fields cause the development of a domain wall interface
along the direction parallel to the surface of the film, as is
shown in Figure \ref{fig2prb}. 
This situation can be described in
terms of a wetting transition that takes place at 
a certain$-$field-dependent$-$critical wetting temperature $T_{w}(h)$. 
In fact, for $T < T_{w}(h)$ a small number of rows parallel to one of
the surface of the film have an overall magnetization pointing to
the same direction as the adjacent surfaces field (see Figure
\ref{fig2prb} (a)). However, the bulk of the film has the opposite
magnetization (i.e. pointing in the direction of the other
competing field). Alternatively, one may also consider a symmetric
situation that is equivalent to the previous one due to the
spin-reversal field-reversal symmetry. This non-wet state of the
surface takes place at low enough temperatures. As the temperature
is raised towards $T_{w}(h)$, the number of rows adjacent to the
surface that has a magnetization of different orientation than the
bulk increases, i.e. the domain wall between the coexisting phase
of opposite magnetization, which at low temperature is tightly
bound to one surface (non-wet state), moves farther and farther
away from the surface toward the bulk of the film. When the
interface is located, on average, in the middle of the film, the
system reaches the wet phase for the first time. Of course, a
well-defined wetting transition takes place in the thermodynamic
limit only. Nevertheless, as shown in Figure \ref{fig2prb}(b), a
precursor of this wetting transition can also be observed in
confined geometries for finite values of $L$. In confined
geometries, this precursor wetting transition is most correctly
described in terms of a localization-delocalization transition of
the interface (for further discussions on the wetting transition
of the Ising, system see e.g.
\cite{abraham80,nakanishi,binder1985,binder1986,abraham1987,privman88,albano1989,
parry1990,parry1992,binder1995,maciolek1996,maciolek1996b,carlon1998,ferrenberg1998,albano2000,
binder2003}). For $T > T_w$, the interface between domains moves
along the $L$-direction, and the system enters the wet regime
(delocalized interface), as is shown in Figure \ref{fig2prb} (c).

The phase diagram  (i.e. the critical curve in the $h-T$ plane, as shown
in Figure \ref{fig3}) has been solved exactly by Abraham \cite{abraham80,abraham1987},
yielding

\begin{equation}
\cosh(2h\beta) = \cosh(2K) - e^{-2K} \sinh(2K) ,
\label{eq:abra}
\end{equation}

\noindent where $J > 0$ is the coupling constant, $h$ is the surface magnetic field,
$\beta = 1/kT$ is the Boltzmann factor, and $K = J\beta$.

\subsubsection{The Corner Geometry and the Filling Transition.}

Another interesting confinement scenario for the Ising magnet is
the so-called corner geometry sketched in Figure \ref{fig2cor}.
For this case, the Hamiltonian given by equation (\ref{eq:ham})
has to be modified for a lattice of size $L \times L$, yielding

\begin{equation}
{\cal H}=-J\sum_{<i,j,m,n>} \sigma_{i,j} \sigma _{m,n} -h \sum_{i}\sigma _{i,1}-
h\sum_{j}\sigma _{L,j}+ h\sum_{j}\sigma _{1,j}+ h \sum_{i}\sigma _{i,L},
\label{eq:hamfil}
\end{equation}

\noindent where $\sigma_{i,j}= \pm 1$ is the spin variable, $J > 0$ is the coupling
constant, and $h$ is the magnitude of the surface field. The first summation runs over
all spins, while the remaining ones hold for spins at the surfaces where the magnetic
fields are applied (see also Figure \ref{fig2cor}) and $h > 0$ is measured in units of $J$.

Studies performed by using this confinement geometry show that for
certain values of the temperature and the competing magnetic
fields $h$, applied to opposite corners, a corner-filling
transition can be observed (see Figure \ref{snapcorner}). The
study of this filling transition under equilibrium conditions has
recently attracted growing attention
\cite{duxbury89,cheng90,napior92,hauge,lipowski1998,rejmer,
parry1999,parry2000,gleiche2000,rascon2000,parry2000b,bednorz2001,parry2001,parry2001b,
abraham2002,abramacio2002,parry2002,sartori,abraham2003,devirgilis2003,milchev2003,
parry2004,romero2004,romero2005,romero2005a,abraham2005,rejmer2005,rascon2005,giuglia,
muller2005}. Also, the filling transition upon the irreversible
growth of a magnetic system has very recently been studied
\cite{manias}. In both cases, the occurrence of an interface
between magnetic domains of different orientation is due to the
presence of competing fields. The localization-delocalization
transition of the interface in a finite system yields to a true
second-order corner-filling transition in the thermodynamic limit
($L \rightarrow \infty$). The analytical expression of the
equilibrium phase diagram was early conjectured by Parry et al.
\cite{parry2001} and more recently proved rigorously by Abraham
and Maciolek \cite{abramacio2002}, yielding

\begin{equation}
\cosh(2h\beta)=\cosh(2K) - e^{-2K} \sinh^2(2K) .
\label{eq:cor}
\end{equation}

The critical curve obtained by using equation (\ref{eq:cor}) is
shown in Figure \ref{fig3}. It is found that for a given surface
magnetic field, the filling transition takes place at a
temperature lower that of the wetting transition, except of course
for $h=0$, where both curves converge to the Onsager critical
temperature of the Ising model ($T_C$).

In all cases briefly discussed here, one has that the interplay
between confinement, boundary conditions, surface fields and
temperature leads to the occurrence of fluctuating interfaces, so
that the propagation of damage in such systems is expected to
strongly depend on the properties of interfaces. It should also be
noticed that due to the equivalence between the Ising model and
both a lattice gas and a binary alloy, all the physical situations
discussed above have a wider field of application, e.g. for the
study of simple fluids and in the field of condensed matter
physics.

\subsection{Damage Spreading.}

\subsubsection{Strip Geometries.}

Let us start our discussion of DS in confined samples with the case of the strip
geometry ($L \times M$, with $L\ll M$) \cite{rubio2001,rubio2002}.
Close to criticality, this Ising magnet exhibits an interesting boundary effect,
as already discussed in relation to Figures \ref{fig1intro} (a) and \ref{fig1intro} (b).
When an initial perturbation is introduced in these configurations, at the critical point
of the DS transition, one observes a monotonic growth of the damage according to a
power-law behavior (see Figure \ref{damobc}) given by

\begin{equation}
D(t)\propto t^{\eta},
\label{eq:pwlaw}
\end{equation}

\noindent where $\eta$ is the dynamic critical exponent. Results
obtained from the study of DS in these systems are consistent with
the fact that the presence of interfaces between magnetic domains
enhances the propagation of the perturbation, as judged by the
values of the dynamic critical exponents: $\eta^{OBC} =0.471(5)$
and $\eta^{PBC}=0.392(5)$ \cite{rubio2001}. Based on the fact that
$\eta^{OBC} > \eta^{PBC}$, one concludes that, for a given time,
the damaged area of the sample is always bigger in samples with
OBC as compared with samples having PBC. This result can be
explained in terms of the fluctuations in the orientation of the
spins in the region near the interfaces. In fact, around these
regions, which are present in samples with OBC (see figure
\ref{fig1intro} (b)), one has the largest fluctuations that
enhance the propagation of the perturbation. In contrast, inside
of the magnetic monodomains, which characterize the samples
obtained by using PBC (see figure \ref{fig1intro} (a)), the
propagation of the damage slows down.

\begin{figure}
\centerline{{\epsfysize=5in \epsffile{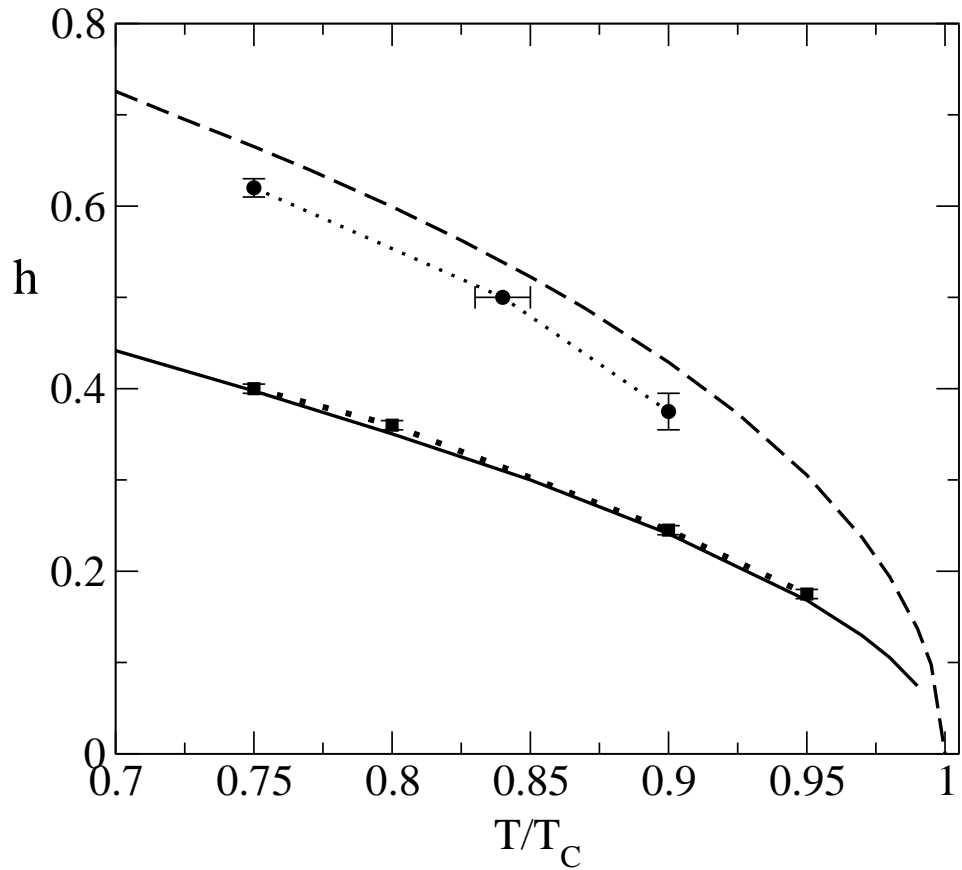}}} 
\caption{Phase
diagram of $h$ versus $T/T_{C}$. The dashed line corresponds to
the wetting transition (equation (\ref{eq:abra})), while the full
line corresponds to the corner-filling transition (equation
(\ref{eq:cor})). The circles and squares correspond to the results
obtained for the DS transition at the strip \cite{rubio2002PRB}
and the corner \cite{rubio2007} geometries, respectively.}
\label{fig3}
\end{figure}

\begin{figure}
\centerline{{\epsfysize=3in \epsffile{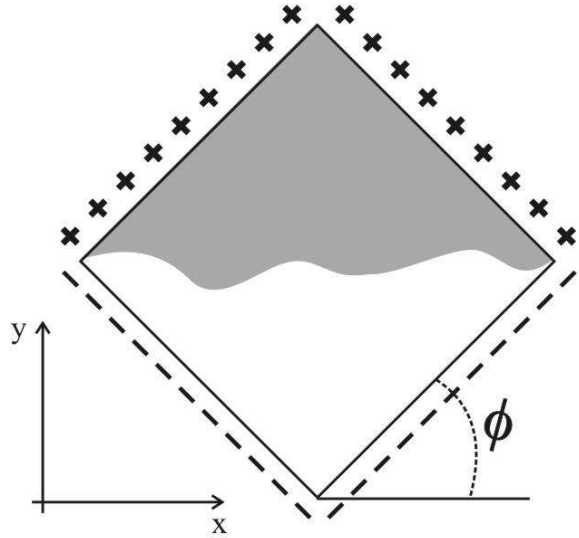}}} 
\caption{Corner
geometry of size $L \times L$. The signs $+$ and $-$ indicate the
surfaces where the competing surface magnetic fields are applied.
The positive domain is shown in grey and the negative one is left
white. In this case, the boundary conditions are open for all
sides of the sample \cite{rubio2007}.}
\label{fig2cor}
\end{figure}

\begin{figure}
\centerline{{\epsfysize=2in \epsffile{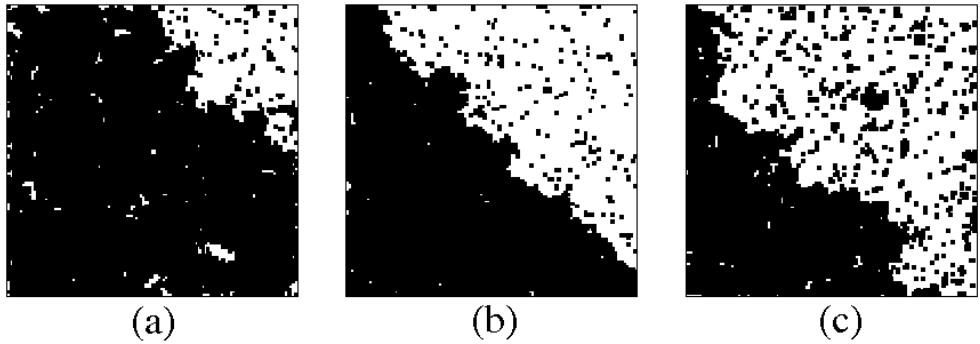}}} 
\caption{Snapshot
configurations obtained after $10^4$ mcs in a corner geometry of
size $L \times L $, with $L = 128$, $T=0.80 T_{C}$ and different
surface fields: (a) $h = 0.20 < h_F(L)$; (b) $h = 0.24 \sim
h_F(L)$, and (c) $h = 0.28 > h_F(L)$. Sites taken by down spins
are shown in black while up spins are left white. More details in
the text.}
\label{snapcorner}
\end{figure}

\begin{figure}
\centerline{{\epsfysize=4in \epsffile{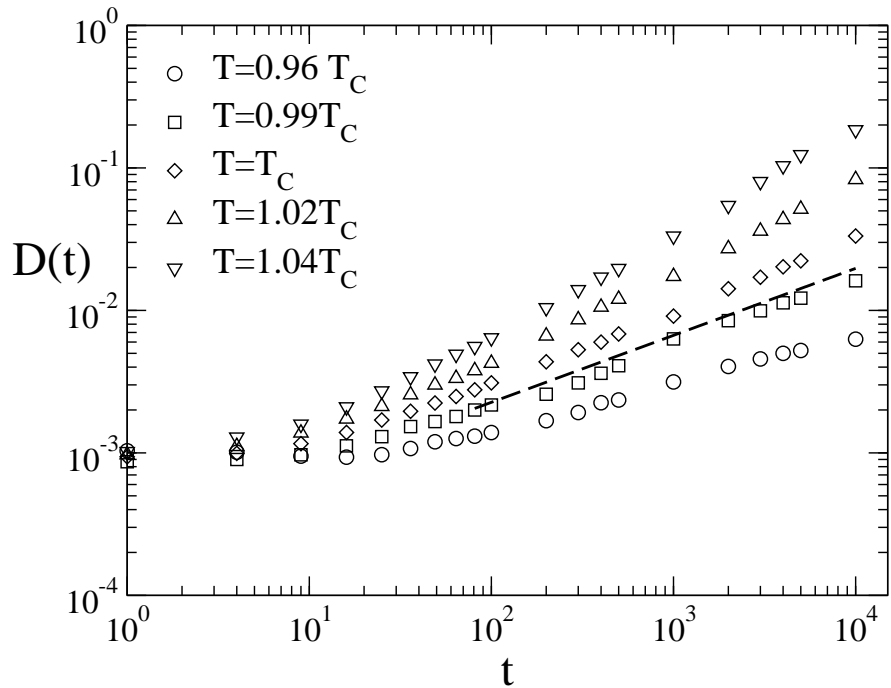}}}
\caption{Log-log plots of $D(t)$ versus $t$ obtained for the Ising magnet in the strip geometry. Results corresponding to different temperatures obtained using lattices of size $L \times M =12\times 601$ and applying open boundary conditions. The slope of the dashed line is $\eta^{OBC}= 0.47$ (see equation \ref{eq:pwlaw}) \cite{rubio2002}. More details in the text.} 
\label{damobc}
\end{figure}

\begin{figure}
\centerline{{\epsfysize=4in \epsffile{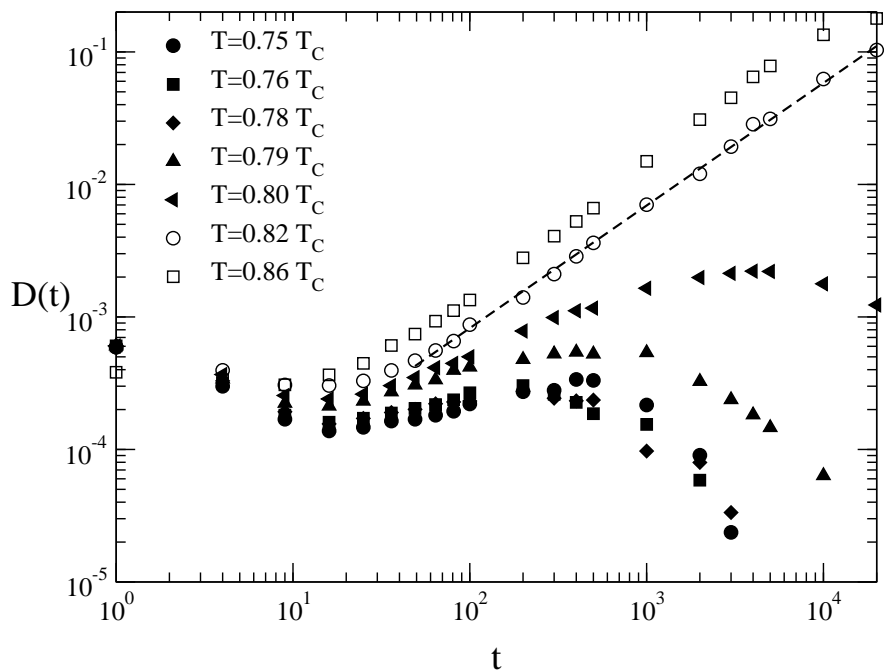}}}
\caption{Log-log
plots of $D(t)$ versus $t$. Results obtained at different
temperatures, applying short-range fields of magnitude $h = 0.5$,
and using lattices of size $L \times M = 24 \times 1201$. The
dashed line has slope $\eta^{WT}= 0.90$ \cite{rubio2002PRB}.}
\label{fig3prb}
\end{figure}

A similar behavior is observed for the propagation of damage in
the Abraham's model (see Figure \ref{fig3prb})
\cite{rubio2002PRB}. In this case, for certain values of the
temperature $T$ and the surface magnetic fields applied to the
upper and lower walls of the lattice ($h$), a critical
localization-delocalization transition of the interface between
magnetic domains occurs. It is found \cite{rubio2002PRB} that the
dynamic critical exponent for DS is bigger than in the previous
cases, yielding $\eta^{WT} = 0.91(1)$ (see Figure \ref{fig3prb}).
This result reflects the fact that the interface between magnetic
domains is parallel to the propagation of the perturbation, and
for this reason one has $\eta^{WT} > \eta^{OBC} > \eta^{PBC}$.
Moreover, after proper extrapolation to the thermodynamic limit,
the phase diagram of the DS transition can be drawn, as shown in
Figure \ref{fig3}. It is observed that the DS transition occurs
within the non-wet phase of the wetting phase diagram (see Figure
\ref{fig3}) and consequently it does not coincide with the wetting
transition.

Other interesting result related to the propagation of damage along the interface between magnetic domains can be visualized by the study of the damage profiles along the $L$-direction, defined as:
\begin{equation}
P_{L}(j,t)=\frac{1}{2M}\sum ^{M}_{i=1}\left| S^{A}_{i,j}(t,T)-S^{B}_{i,j}(t,T)]\right|,
\label{eq:perfil}
\end{equation}

\noindent where  $S^{A}_{i,j}(t,T)$ and $S^{B}_{i,j}(t,T)$ are the reference and damaged
configurations at site $l$ of coordinates $\{i,j\}$, as early defined in the context of
equation (1), respectively.  Thus, the damage profile represents the average damage 
of the $i-th$ ($i = 1,....,L$) row of the system, that runs parallel to the 
surfaces where the fields are applied, i.e. the $M$-direction.

\begin{figure}
\centerline{\epsfig{file=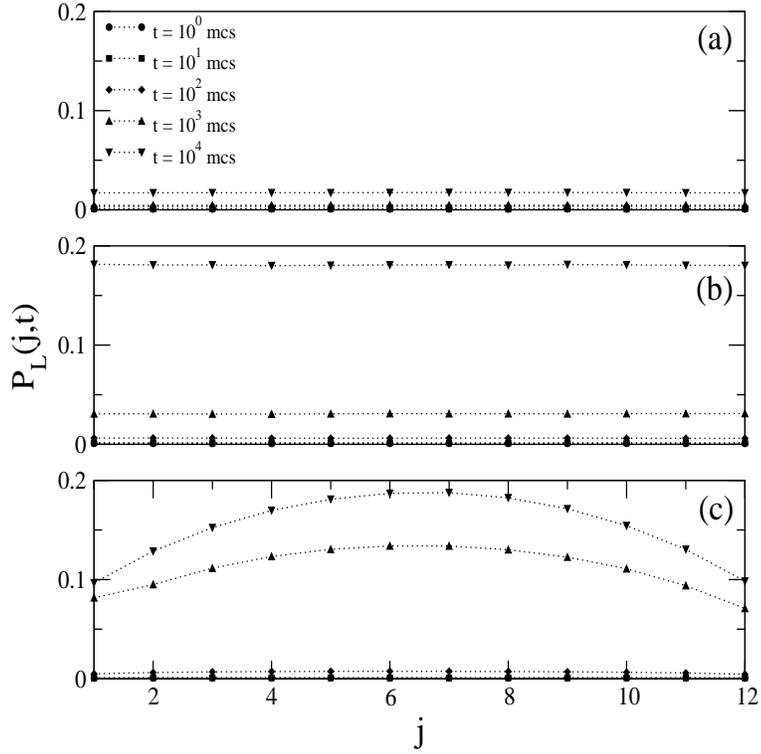,height=14cm,width=12cm,clip=,angle=-90}}
\caption{Plots of the damage profile measured along the L-direction and 
obtained at different times as indicated. Results are obtained using lattice of size $L=12$, $M=600$ in confined geometries, namely (a) $T=0.98 T_{C}$ 
and PBC, (b) $T=0.98 T_{C}$ and OBC, (c) $T=0.861 T_{C}$ and applying short-range fields $\left| h_{1}\right| =\left| h_{L}\right| =0.5$.\cite{rubio2002} }
\label{perfil}
\end{figure}

It is found \cite{rubio2002} that for zero fields and assuming both PBC and OBC 
(Figure \ref{perfil} (a)-(b), respectively), the damage profiles are essentially 
flat. This can be explained in terms of the interfaces between different magnetic domains which are essentially perpendicular 
to the film and the overall effect on the profiles is homogeneous.
On the other hand, in presence of surface fields the situation is qualitatively 
different (see Figure \ref{perfil}(c)). The damage profiles show an important curvature 
related to the existence of an interface in the $M$-direction, as shown in Figure \ref{fig2prb}(c). The effect of the walls is to slow down the propagation of the damage, while the interface between magnetic domains running along the film enhance the spreading of the damage.

\subsubsection{Corner Geometry.}

\begin{figure}
\centerline{{\epsfysize=4in \epsffile{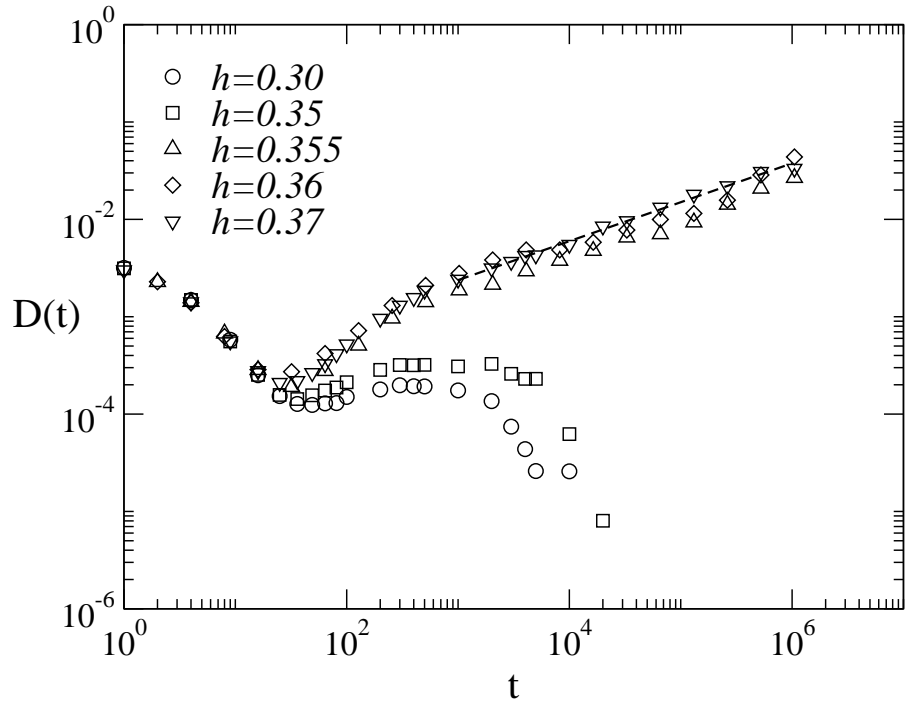}}}
\caption{Log-log plots of $D(t)$ versus $t$, obtained for the Ising magnet in the corner geometry and for different values of the surface magnetic field $h$. Results corresponding to $T=0.80 T_C$ and obtained using lattices of size $L = 256$. The dashed line has slope $\eta^{**}= 0.40$ \cite{rubio2007}. More details in the text.}
\label{fig4cor}
\end{figure}

\begin{figure}
\centerline{{\epsfysize=4in \epsffile{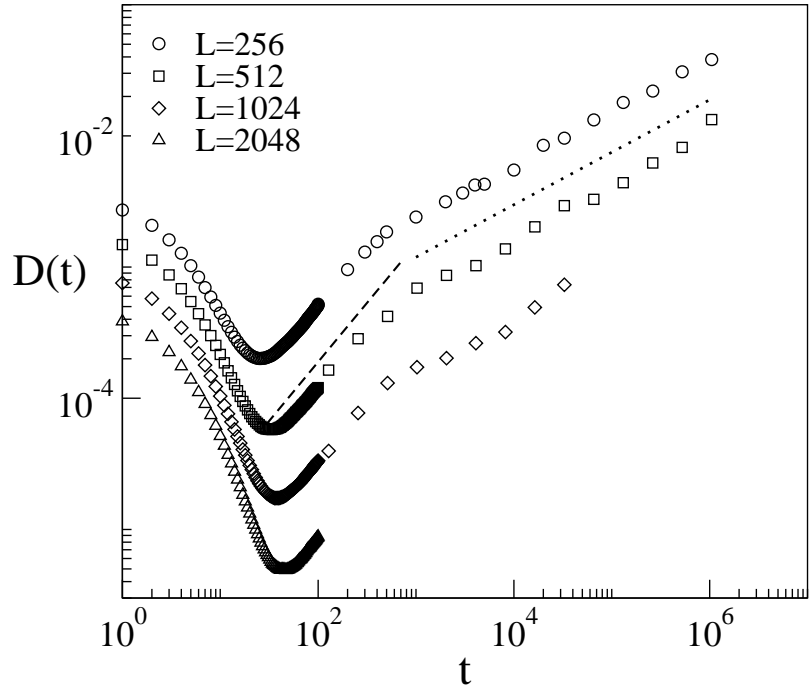}}} 
\caption{Log-log
plot of $D(t)$ versus $t$, as obtained for the Ising magnet in the
corner geometry. Results obtained for $T=0.80 T_C$ and at the
``critical'' size-dependent magnetic field $h_D(L)$:
$h_D(L=256)=0.355$, $h_D(L=512)=0.3675$, $h_D(L=1024)=0.37$, and
$h_D(L=2048)=0.38$. The dashed line has slope $\eta^{*} = 0.89$
and the dotted line has slope $\eta^{**} = 0.40$ \cite{rubio2007}.} 
\label{fig5cor}
\end{figure}

DS in the corner geometry, with competing (short-range) magnetic
fields acting on the surfaces, shows \cite{rubio2007} a behavior
qualitatively similar to that the observed in the case of the
strip geometry (see Figure \ref{fig4cor}). However, in contrast to
the previous case, three different regimes for DS were found, as
is shown in Figure \ref{fig5cor}. For short times, one observes
the healing of the damage created in the bulk of the domains.
However, at the critical DS point, a small cluster of damaged
sites survives close to the interface. During the second time
regime, the damage propagates along the interface according to a
power-law behavior of the form $D(t) \propto t^{\eta^*}$, with
$\eta^{*} = 0.89(1)$. Finally, due to the constraint imposed by
the corners where magnetic fields of opposite direction meet, the
damage no longer propagates along the surface but starts to spread
slowly into the bulk of the domains. Within this late regime one
has $D(t) \propto t^{\eta^{**}}$, where $\eta^{**} = 0.40(2)$ is the
exponent describing the spreading of the damage in the bulk
\cite{rubio2007}.

\begin{figure}
\centerline{{\epsfysize=4in \epsffile{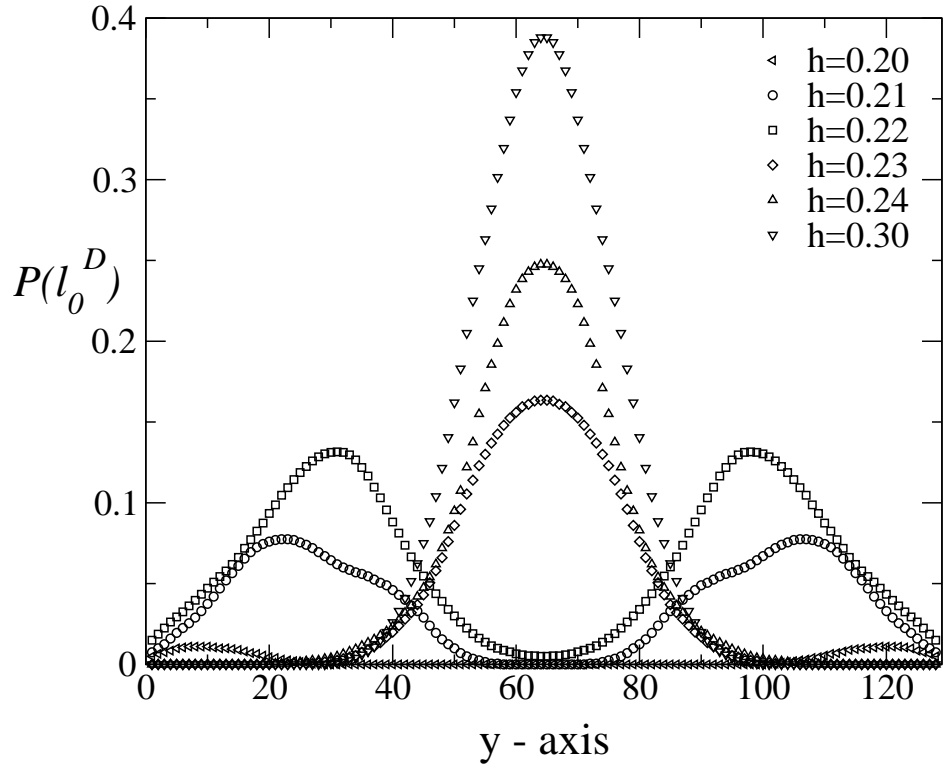}}} 
\caption{Plot of probability distribution of the position of the damage $P(l_0^D)$, obtained for $L=128$, $T=0.90 T_C$, and different values of surface magnetic field $h$, as listed in the figure \cite{rubio2007}.} 
\label{ploD}
\end{figure}

As in the case of strip geometry, it is useful to gain further insight into the spatiotemporal propagation of 
the damage. For this purpose, the probability distribution of the 
 distance from the damage zone to the corner ($P(l_0^{D})$) can be measured.
The distribution was evaluated along the diagonal of the sample ($y$-direction in Figure \ref{fig2cor}) at $x=L/2$. 
Figure \ref{ploD} shows a summary of the results \cite{rubio2007}. 

In these cases, for $h\ll h_f(\infty)$ (e.g. $h=0.20$) the distribution is almost flat with 
two small peaks close to the corners. 
Approaching the transition by increasing the field these
 peaks develop and become slightly shifted toward the center of the sample 
(e.g. $h=0.21$ and $h=0.22$ in Figure \ref{ploD}).
This double-peaked structure indicates that the 
damage remains bound to each corner with the same probability as expected for the case of the non-wet phase.
On the other hand, for $h\sim h_f(\infty)$ (e.g. $h=0.23, 0.24$ in Figure \ref{ploD}) the distribution becomes a Gaussian centered along the middle of the sample. 
The Gaussian structure of $P(l_0^{D})$ remains even for $h\gg h_f(\infty)$ 
(e.g. $h=0.30$ in Figure \ref{ploD}).
This results are in qualitative agreement with the corresponding to the damage profile showed in Figure \ref{perfil} for the strip geometry; and it shows that the damage is located in the neighborhood of the interface between magnetic domains.

\subsubsection{Overview of the Behaviour of DS in Different Confinement Geometries.}

Table I summarizes the exponents obtained for different
confinement geometries and boundary conditions. In view of the
exponents obtained, one concludes that the presence of interfaces
between magnetic domains of different orientation causes the
enhancement of the propagation as compared with samples where such
interfaces are not well defined. Furthermore, these results
clearly show that the propagation of the damage is anisotropic: it
propagates better along an interface than across it. Moreover, the
obtained results suggest that the damage propagates along the
magnetic interfaces with a critical exponent $\eta^{I} = 0.90(2)
\simeq \eta^{WT} \simeq \eta^{*}$ and it spreads into the magnetic
domains with an exponent $\eta^{DOM} = 0.40(2) \simeq \eta^{PBC}
\simeq \eta^{**}$.

Taking into account this point of view, the value of the exponent
$\eta^{OBC} =0.471(5)$ can be explained in terms of the
equilibrium configurations of the system. In fact, the sequence of
domains with spins of opposite sign implies the inhomogeneity in
the propagation of damage. When the perturbation arrives at an
interface, it will propagate faster than in the bulk. For this
reason, the exponent $\eta^{OBC}$ can be thought as a result of
the interplay between the propagation along the interfaces and the
propagation inside the magnetic domains. Regrettably, for this
example it seems to be impossible to separate both effects in
order to write the exponent in terms of $\eta^{I}$ and
$\eta^{DOM}$.

\begin{figure}
\centerline{{\epsfysize=4in \epsffile{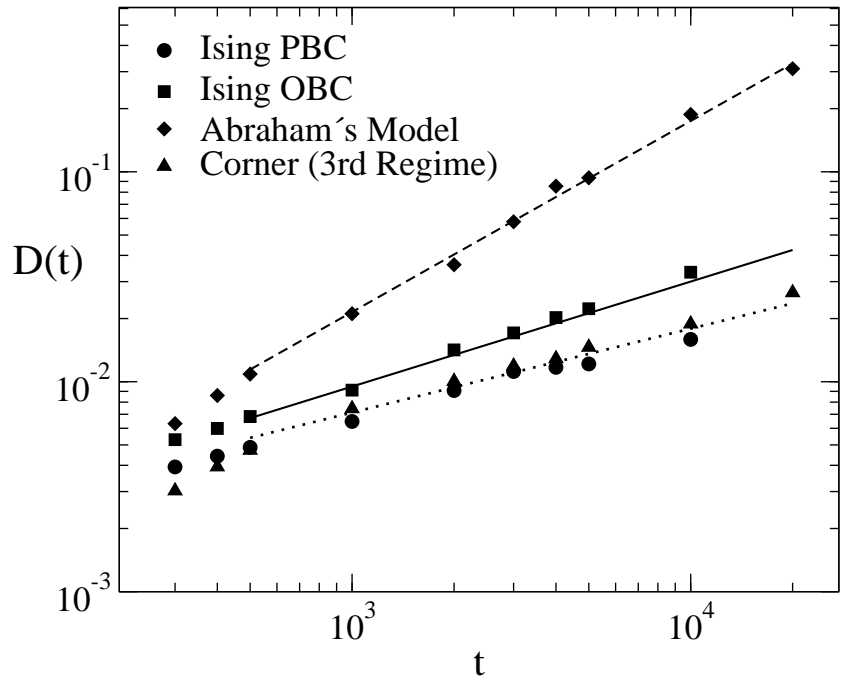}}} 
\caption{Log-log
plot of $D(t)$ versus $t$ obtained at the DS critical point for
the different models described in this work. The dotted line has
slope $\eta^{DOM} = 0.40$, the dashed line has slope $\eta^{OBC} =
0.47$, and the solid line has slope $\eta^{I} = 0.90$. More
details in the text.} 
\label{pendientedano}
\end{figure}

Figure \ref{pendientedano} shows the behavior of $D(t)$ for the
different confinement geometries studied: the strip geometry with
PBC's and OBC's at criticality and without applied magnetic fields
\cite{rubio2001, rubio2002}, the Abraham's Model
\cite{rubio2002PRB} and the corner geometry \cite{rubio2007}. In
all these cases and because of the dependence on the initial
conditions, the first $100$ mcs were disregarded in order to
evaluate the exponents.

\vspace{0.3cm}
{\centering \begin{tabular}{|c|c|}
\hline
Model&
Critical Exponent $\eta$ \\
\hline
\hline
Ising strip geometry with PBC \cite{rubio2001}&
$0.392(5)$\\
\hline
Ising strip geometry with OBC \cite{rubio2001}&
$0.471(5)$\\
\hline
Abraham's Model \cite{rubio2002PRB}&
$0.91(1)$\\
\hline
Corner Geometry (Second Regime) \cite{rubio2007}&
$0.89(1)$\\
\hline
Corner Geometry (Third Regime) \cite{rubio2007}&
$0.40(2)$\\
\hline
\end{tabular}\par}
\vspace{0.3cm}

{\bf Table I.} List of the critical exponent of Damage Spreading ($\eta$)
obtained for the different models, as indicated in the first column.

\section{Conclusions.}

A brief overview of the Damage Spreading method applied to
different models is presented and the critical non-equilibrium DS
transition is analyzed. Within this context, DS in confined
geometries is discussed in detail. A power-law behavior is found
for DS as a function of time. The dynamic critical exponent $\eta$
depends on the presence of interfaces between magnetic domains
generated by the existence of surface magnetic fields applied to
the walls of the lattice. The results reported suggest that the
damage propagates along the interfaces between domains with
opposite orientations of the magnetization with an exponent
$\eta^I\sim 0.90$ and it spreads into the magnetic domain with an
exponent $\eta ^{DOM}\sim 0.40$. In view of these results, we
conclude that the presence of interfaces enhances the propagation
of a perturbation.

The critical propagation-nonpropagation transition of the damage
was studied too. The results discussed indicate that, in some
cases as in the Abraham's model and a pure Ising system in the
absence of magnetic fields, there is clear evidence indicating
that the DS transition and the critical transition of the
corresponding system occur at different critical points. These
cases correspond to the wetting and the ferromagnet-paramagnet
transitions, respectively. In the case of the corner geometry,
both transitions $-$damage spreading and corner-filling$-$ coincide
within error bars.

In the cases of confined geometries, the critical exponents found
suggest that the DS transition does not belong to the DP
universality class. This finding seems to be related to the
enhancement of the propagation of the damage caused by the
magnetic interfaces appearing in confined samples.

The DS method is a suitable tool for the study and understanding
of the propagation of perturbations in physical systems. Due to
the concept of universality that allows one to describe
second-order phase transitions and classify them into universality
classes, characterized by the same set of critical exponents, it
is possible to investigate simple models representing more complex
physical situations, as e.g. the Ising model as archetype of
various systems such as ferromagnets, fluids, lattice gases,
binary alloys, etc. Therefore, the discussed role of the
interfaces in the propagation of perturbations may be a quite
general phenomenon of relevance in material science in general,
and particularly, due to the additional influence of confinement,
in the field of micro- and nanotechnology.

\vskip 1.0 true cm
\noindent{\bf  Acknowledgments}. This work was supported financially by
CONICET, UNLP, and ANPCyT (Argentina). 
MLRP acknowledges CONICET for the grant of a fellowship.

\newpage

\newpage

\newpage

\newpage

\newpage

\newpage

\newpage

\newpage

\newpage

\newpage

\newpage

\newpage

\end{document}